\documentclass[aps,pra,preprint,superscriptaddress,longbibliography]{revtex4-2}

\usepackage{amsmath}
\usepackage{amsthm}
\usepackage{amsfonts}
\usepackage{amssymb}
\usepackage{xcolor,graphicx}
\usepackage{tikz}
\usepackage{color}
\usepackage[pdftex]{hyperref} 
\usepackage{longtable}
\usepackage{array}
\usepackage{dsfont}

\begin{document}

	\title{A curvilinear framework for vector light fields}

    \author{Leonardo S. Gonzalez-Aceves}
	\affiliation{Photonics and Mathematical Optics Group, Tecnologico de Monterrey, Monterrey 64849, Mexico.}

    \author{Gabriela Flores-Cova}
	\affiliation{Centro de Investigaciones en Óptica, A.C., Loma del Bosque 115, Colonia Lomas del campestre, 371507 León, Gto., Mexico.}

    \author{B.~M. Rodr\'iguez-Lara}
	\email[e-mail: ]{bmlara@upp.edu.mx; 
    blas.rodriguez@gmail.com}
	\affiliation{Universidad Polit\'ecnica de Pachuca. Carr. Pachuca-Cd. Sahag\'un Km.20, Ex-Hda. Santa B\'arbara. Zempoala, 43830 Hidalgo, M\'exico }

    \author{Raul I. Hernandez-Aranda}
	\affiliation{Photonics and Mathematical Optics Group, Tecnologico de Monterrey, Monterrey 64849, Mexico.}

    \author{Alfonso Jaimes-Najera}
	\affiliation{Photonics and Mathematical Optics Group, Tecnologico de Monterrey, Monterrey 64849, Mexico.}

    \author{Carmelo Rosales-Guzman}
	\affiliation{Centro de Investigaciones en Óptica, A.C., Loma del Bosque 115, Colonia Lomas del campestre, 371507 León, Gto., Mexico.}

    \author{Benjamin Perez-Garcia}
	\email[e-mail: ]{b.pegar@tec.mx}
	\affiliation{Photonics and Mathematical Optics Group, Tecnologico de Monterrey, Monterrey 64849, Mexico.}

	\date{\today}
	
	\begin{abstract}
        Vector beams are often regarded as non-separable superpositions of spatial and polarization degrees of freedom that satisfy the wave equation.
        This interpretation ties their polarization structure to their spatial shape.
        Here, we introduce a generalized method to construct vector beams whose structure is entirely encoded in the polarization degree of freedom.
        Using conformal maps, we construct orthonormal polarization bases from the geometry of the coordinates and encode them experimentally via phase-only spatial light modulators. 
        We apply our method to four systems, elliptical, parabolic, bipolar, and dipole, that represent algebraic and transcendental families of conformal maps. 
        Stokes polarimetry measurements confirm agreement with theoretical predictions.
	\end{abstract}
	
	\maketitle
	\newpage


\section{Introduction}

The ability to control light across multiple degrees of freedom (DoFs), including phase, amplitude, polarization, and wavelength, has reshaped the landscape of photonic applications, from high-resolution microscopy to quantum cryptography~\cite{Roadmap2017}. 
Among these structured light fields, vector beams, defined as non-separable superpositions of spatial and polarization DoFs, have enabled a wide range of applications~\cite{RosalesGuzman2024b} and motivated concepts such as classical entanglement~\cite{Forbes2019} and optical Skyrmions~\cite{Shen2024}. 
This has increased the demand for techniques to generate and characterize vector beams~\cite{RosalesGuzman2018,Shen2022}.

The spatial DoF of light can be tailored using optical elements such as axicons, cylindrical lenses, spiral phase plates, and diffractive optics. 
Recent advances have introduced computer-controlled devices, among which liquid crystal spatial light modulators (SLMs) and digital micromirror devices (DMDs) are the most widely used~\cite{RosalesGuzman2024,Hu2022,Scholes2019}. 
Both enable phase- or amplitude-only modulation of complex light fields, but their polarization response differs. 
SLMs modulate only linearly polarized light, typically in the horizontal state. 
Generating vector beams with an SLM requires separate control of spatial and polarization DoFs, often implemented through interferometric arrays~\cite{Maurer2007,Niziev2006} or cascaded sections of a single device~\cite{Rong2014,Otte2018b,RodriguezFajardo2024}.
In contrast, DMDs are polarization-insensitive and support direct modulation of arbitrary polarization states~\cite{Rosales2020}.

Although these devices can produce light fields with arbitrary phase and amplitude distributions, most implementations adopt exact or paraxial solutions to the wave equations due to their analytical tractability and experimental relevance.

Exact solutions include Bessel~\cite{Jauregui2005}, Mathieu~\cite{RodriguezLara2008}, and Weber~\cite{RodriguezLara2009} fields, while paraxial solutions include Hermite-, Laguerre-, and Ince-Gauss beams~\cite{Siegman1986,Bandres2004}. 
Each family forms an infinite set of orthogonal modes and serves as a foundation for structured beam design. 
These families have reinforced the view that vector beams should exhibit spatial distributions confined to predefined modal bases. 
Typical implementations rely on Bessel fields~\cite{Jauregui2005,Dudley2013} and Laguerre-Gauss beams~\cite{Galvez2012} to impose polar symmetry; Mathieu fields~\cite{RodriguezLara2008,Rosales2021Mathieu} and Ince-Gauss beams~\cite{Liyao2020} to realize elliptical symmetry; and more recent constructions based on parabolic~\cite{RodriguezLara2009,ZhaoBo2022}, helico-conical~\cite{MedinaSegura2023}, and Pearcey~\cite{RodriguezFajardo2024Pearcey} beams to produce intricate spatial profiles.

Current approaches constrain vector beams to predefined spatial mode families, which limits control over their spatial and polarization structure. 
We introduce a method to construct an orthonormal polarization basis adapted to curvilinear coordinates, removing this constraint. 
Our approach yields closed-form expressions suitable for computer-generated holograms on phase-only SLMs, enabling the experimental generation of vector beams with spatially varying polarization distributions beyond conventional modal families. 
Certain curvilinear coordinate systems additionally allow tunable control over the degree of polarization by adjusting geometric parameters, offering a direct means to manipulate the beam’s vectorial character.

Our manuscript is organized as follows. 
In Sec.~\ref{sec:Sec2}, we introduce our framework for constructing curvilinear polarization bases from coordinate systems defined by conformal maps. 
We apply our framework in Sec.~\ref{sec:Sec3} to four representative geometries, elliptical, parabolic, bipolar, and dipole, and derive the corresponding polarization structures. 
In Sec.~\ref{sec:Sec4}, we describe our experimental setup for implementing these structures with phase-only spatial light modulators. 
We present our reconstructed polarization distributions, obtained via Stokes polarimetry, and compare them with our theoretical predictions in Sec.~\ref{sec:Sec5}. 
We conclude in Sec.~\ref{sec:Sec6} by summarizing our results and outlining future directions for extending our approach.

\section{Theoretical Framework} \label{sec:Sec2}

We describe a coherent light field as a linear superposition of orthonormal spatial and polarization modes,
\begin{align}
    \mathbf{E}(\mathbf{r}) = \int d\boldsymbol{\alpha} ~ \sum_{u = 1, 2} c_{\mathbf{k},u} \psi_{\alpha}(\mathbf{r}) \hat{e}_{u}(\mathbf{r}),
\end{align}
where $\psi_{\alpha}(\mathbf{r})$ and $\hat{e}_{u}(\mathbf{r})$ are elements of orthonormal spatial and polarization bases, respectively. 
Plane waves, $\psi_{\mathbf{k}}(\mathbf{r}) = e^{i \mathbf{k} \cdot \mathbf{r}}$, combined with the horizontal and vertical linear polarization states, $\hat{e}_{x}$ and $\hat{e}_{y}$, form a simple example. 
More structured bases include Bessel waves, $\psi_{k, k_{\perp}, m}(\mathbf{r}) = J_{m}(k_\perp \rho) e^{i m \phi} e^{i k_z z}$~\cite{Jauregui2005}, and radial or azimuthal polarization, defined by $\hat{e}_{\rho}(\mathbf{r}) = \cos \phi \, \hat{e}_{x} + \sin \phi \, \hat{e}_{y}$ and $\hat{e}_{\phi}(\mathbf{r}) = -\sin \phi \, \hat{e}_{x} + \cos \phi \, \hat{e}_{y}$~\cite{Tidwell1990}. 
While Cartesian and cylindrical bases suffice in many cases, more specialized applications require polarization bases adapted to spatial symmetries. 
We use curvilinear coordinates to construct such bases directly.

Structured light fields are often derived by solving the Helmholtz equation in curvilinear coordinates~\cite{Jauregui2005,RodriguezLara2008,RodriguezLara2009}, anchoring spatial and polarization DoFs to the geometry~\cite{VolkeSepulveda2006}. 
In our approach, we construct orthonormal polarization bases directly from general curvilinear coordinates. 
Given a coordinate map $f: (u,v) \rightarrow (x,y)$ in the transverse plane, we define the orthonormal unit vectors,
\begin{align}
    \begin{aligned}
        \hat{e}_{u} &= \frac{1}{h_{u}} \frac{\partial \mathbf{r}}{\partial u} =  \frac{1}{h_{u}} \left( \frac{\partial x}{\partial u} \hat{e}_{x} +  \frac{\partial y}{\partial u} \hat{e}_{y} \right), \\
        \hat{e}_{v} &= \frac{1}{h_{v}} \frac{\partial \mathbf{r}}{\partial v} = \frac{1}{h_{v}} \left( \frac{\partial x}{\partial v} \hat{e}_{x} +  \frac{\partial y}{\partial v} \hat{e}_{y} \right),
    \end{aligned}
\end{align}
in terms of the scaling factors,
\begin{align}
    \begin{aligned}
        h_{u} &= \left\| \frac{\partial \mathbf{r}}{\partial u} \right\| = \sqrt{ \left( \frac{\partial x}{\partial u} \right)^{2} + \left( \frac{\partial y}{\partial u} \right)^{2}}, \\
        h_{v} &= \left\| \frac{\partial \mathbf{r}}{\partial v} \right\| = \sqrt{ \left( \frac{\partial x}{\partial v} \right)^{2} + \left( \frac{\partial y}{\partial v} \right)^{2}}, 
    \end{aligned}
\end{align}
which are the norms of the partial derivatives of the position vector $\mathbf{r} = x \hat{e}_{x} + y \hat{e}_{y} + z \hat{e}_{z}$~\cite{Arfken2011}. 
This geometric construction allows us to generate and manipulate polarization structures without relying on a predefined spatial mode basis.

We require the coordinate map $f: (u, v) \to (x, y)$ to be locally invertible, which guarantees the existence of an inverse map $f^{-1}: (x, y) \to (u, v)$ within a neighborhood of each point. 
Local invertibility ensures that our polarization states remain physically meaningful and experimentally accessible across the transverse plane. 
However, most coordinate maps are not globally one-to-one, since multiple points in the domain can correspond to the same coordinates in the range. 
A familiar example is the polar transformation $f: (x, y) \to (\rho, \phi)$, defined by $\rho = \sqrt{x^2 + y^2}$ and $\phi = \arctan(y/x)$. 
This map is locally invertible, and we can recover Cartesian coordinates via $f^{-1}: (\rho, \phi) \to (x, y)$, with $x = \rho \cos \phi$ and $y = \rho \sin \phi$. 
Still, the periodicity of $\phi$ prevents global invertibility, since multiple polar pairs $(\rho, \phi)$ correspond to a single Cartesian pair $(x, y)$. 
To avoid ambiguity, we typically restrict the range to $\rho \in [0, \infty)$ and $\phi \in [0, 2\pi)$. 
Even with this restriction, the origin $\rho = 0$ remains a singular point where $\phi$ is undefined. 
Rather than obstructing our analysis, this singularity highlights the importance of local geometric structure in our framework.

In typical experimental setups, SLMs introduce a spatially varying phase modulation onto a single linear polarization component, usually horizontal. 
For example, an input field with diagonal polarization, $\hat{e}_{\mathrm{in}} = (\hat{e}_{x} + \hat{e}_{y})/\sqrt{2}$, emerges from the SLM as $\hat{e}_{\mathrm{out}} = ( e^{i \varphi} \hat{e}_{x} + \hat{e}_{y})/\sqrt{2}$. 
Placing a quarter-wave plate (QWP) after the SLM with its fast axis oriented at $45^\circ$ transfers the phase modulation onto a circular polarization component. 
In this arrangement, an input state $\hat{e}_{\mathrm{in}} = (\hat{e}_{L} + \hat{e}_{R})/\sqrt{2}$ becomes $\hat{e}_{\mathrm{out}} = (\hat{e}_{L} + e^{i \varphi} \hat{e}_{R})/\sqrt{2}$. 
Motivated by this configuration, we adopt the circular polarization basis to simplify the implementation of our curvilinear polarization vectors and establish a direct link between theory and experiment.

In the circular polarization basis, we express the curvilinear unit vectors,
\begin{align}
    \begin{aligned}
        \hat{e}_{u}(\mathbf{r}) &= \frac{1}{\sqrt{2} h_{u}} \left( \frac{\partial z^{\ast}}{\partial u} \hat{e}_{L} + \frac{\partial z}{\partial u} \hat{e}_{R} \right), \\
        \hat{e}_{v}(\mathbf{r}) &= \frac{1}{\sqrt{2} h_{v}} \left( \frac{\partial z^{\ast}}{\partial v} \hat{e}_{L} + \frac{\partial z}{\partial v} \hat{e}_{R} \right),
    \end{aligned}
\end{align}
using the complex representation $z = x + i y$. 
This form naturally leads us to conformal maps, defined as complex differentiable transformations $w = f(z)$ with $w = u + i v$ and $f^{\prime}(z) \neq 0$ in the domain of interest. 
Working with conformal maps simplifies the derivation of the partial derivatives,
\begin{align}
    \begin{aligned}
        \frac{\partial z}{\partial u} &= h_{u} e^{i \theta_{u}(\mathbf{r})}, \\
        \frac{\partial z}{\partial v} &= h_{v} e^{i \theta_{v}(\mathbf{r})},
    \end{aligned}
\end{align}
with the phases,
\begin{align}
    \begin{aligned}
        \theta_{u}(\mathbf{r}) &= \arctan \left(\frac{\partial y / \partial u}{\partial x / \partial u} \right), \\
        \theta_{v}(\mathbf{r}) &= \arctan \left(\frac{\partial y / \partial v}{\partial x / \partial v} \right).
    \end{aligned}
\end{align}
Restricting our framework to conformal coordinates~\cite{Kythe2019} allows us to invoke the Cauchy-Riemann equations~\cite{Wunsch1983},
\begin{align}
    \begin{aligned}
        \frac{\partial u}{\partial x} &= \frac{\partial v}{\partial y}, \quad  
        \frac{\partial u}{\partial y} = -\frac{\partial v}{\partial x}, \\
        \frac{\partial x}{\partial u} &= \frac{\partial y}{\partial v}, \quad  
        \frac{\partial y}{\partial u} = -\frac{\partial x}{\partial v},
    \end{aligned} \label{eq:d2zdudv}
\end{align}
from which we obtain the relation,
\begin{align}
    \frac{\partial z}{\partial v} = i \frac{\partial z}{\partial u},
\end{align}
which allows us to recognize a constant offset $\theta_{v} - \theta_{u} = \pi / 2$ between the phases.
Substituting into our expression for the unit vectors, we obtain, up to an overall phase,
\begin{align}
    \begin{aligned}
        \hat{e}_{u}(\mathbf{r}) &= \frac{1}{\sqrt{2}} \left( \hat{e}_{L} + e^{i 2 \theta_{u}(\mathbf{r})} \hat{e}_{R} \right), \\
        \hat{e}_{v}(\mathbf{r}) &= \frac{1}{\sqrt{2}} \left( \hat{e}_{L} - e^{i 2 \theta_{u}(\mathbf{r})} \hat{e}_{R} \right),
    \end{aligned} \label{eq:UnitVectors}
\end{align}
where both vectors depend on a single spatially varying phase $\theta_{u}(\mathbf{r})$, which we can implement directly using an SLM.

We characterize our curvilinear polarization bases using the Stokes parameters. For the unit vector $\hat{e}_{u}$,
\begin{align}
    \begin{aligned}
        S_{0}(\hat{e}_{u}) &= 1, \\
        S_{1}(\hat{e}_{u}) &= \cos 2 \theta_{u}(\mathbf{r}), \\
        S_{2}(\hat{e}_{u}) &= \sin 2 \theta_{u}(\mathbf{r}), \\
        S_{3}(\hat{e}_{u}) &= 0,
    \end{aligned}
\end{align}
while for $\hat{e}_{v}$,
\begin{align}
    \begin{aligned}
        S_{0}(\hat{e}_{v}) &= 1, \\
        S_{1}(\hat{e}_{v}) &= -\cos 2 \theta_{u}(\mathbf{r}), \\
        S_{2}(\hat{e}_{v}) &= -\sin 2 \theta_{u}(\mathbf{r}), \\
        S_{3}(\hat{e}_{v}) &= 0.
    \end{aligned}
\end{align}
These distributions describe spatially varying linear polarization across the transverse plane. 
Introducing the complex Stokes field ~\cite{Freund2001,Dennis2002}, 
\begin{align}
    S = S_{1} + i S_{2}, 
\end{align}
we realize that the Stokes fields for the unit vectors,
\begin{align}
    \begin{aligned}
        S(\hat{e}_{u}) &= e^{i 2 \theta_{u}(\mathbf{r})}, \\
        S(\hat{e}_{v}) &= -e^{i 2 \theta_{u}(\mathbf{r})} = e^{i 2 \theta_{v}(\mathbf{r})},
    \end{aligned} \label{eq:StokesFields}
\end{align}
are pure phase distributions governed by the relation, $\theta_{v}(\mathbf{r}) = \theta_{u}(\mathbf{r}) + \pi/2$.

We express the degree of polarization over a region $\Omega$ in the transverse plane,
\begin{align}\label{eq:degree}
    p_{\Omega} = \frac{ \sqrt{\mathbb{S}_1^2 + \mathbb{S}_2^2 + \mathbb{S}_3^2}}{\mathbb{S}_{0}},
\end{align}
in terms of the global Stokes parameters,
\begin{align}
    \mathbb{S}_j = \iint_\Omega S_{j}(\mathbf{r}) \, d^2\mathbf{r},
\end{align}
with $j=0,1,2,3$, which quantify the total polarization content integrated over the region. 
A value of $p_{\Omega} = 1$ indicates that a single polarization state is present across the entire region, while $p_{\Omega} = 0$ corresponds to a spatially varying polarization distribution that averages to zero net polarization. 
For the curvilinear polarization bases aligned with $\hat{e}_{u}$ and $\hat{e}_{v}$, the degree simplifies to
\begin{align}
    p_{\Omega}(\hat{e}_{u}) = p_{\Omega}(\hat{e}_{v}) =  \frac{ \vert \mathbb{S}_{1}(\hat{e}_{u}) + i \mathbb{S}_{2}(\hat{e}_{u}) \vert}{\mathbb{S}_{0}},
\end{align}
as a consequence of the local structure imposed by the conformal map $z = f(w)$, with $z = x + i y$ and $w = u + i v$. 
Using Eqs.~(\ref{eq:d2zdudv}) and~(\ref{eq:StokesFields}), we obtain
\begin{align}
    \begin{aligned}
        \mathbb{S}_{0} &= \iint_{\Omega} dx\,dy, \\
        \left| \mathbb{S}_{1}(\hat{e}_{u}) + i \mathbb{S}_{2}(\hat{e}_{u}) \right| &= \left| \iint_{\Omega} du\,dv \left( \frac{d f(w)}{dw} \right)^2 \right|,
    \end{aligned}
\end{align}
which can be evaluated analytically in the natural domain of each coordinate system.

\section{Examples in Selected Curvilinear Geometries} \label{sec:Sec3}
\begin{figure}[ht]
    \centering
    \includegraphics[width= 0.75 \columnwidth]{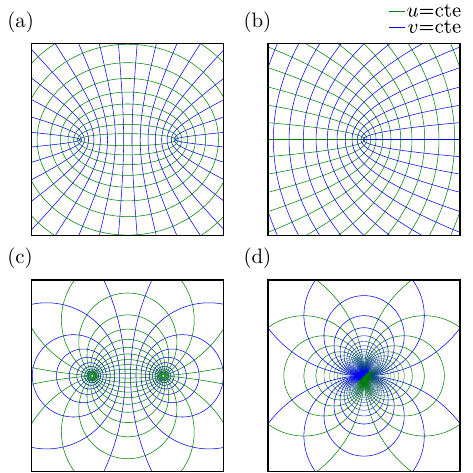}
    \caption{Curvilinear coordinate systems considered: 
    (a) elliptical coordinates, showing confocal ellipses and hyperbolas; 
    (b) parabolic coordinates, with orthogonal parabolas opening in opposite directions; 
    (c) bipolar coordinates, with intersecting and non-intersecting circular arcs; and 
    (d) dipole coordinates, defined by orthogonal families of non-concentric circles.
    }
    \label{fig:curvilinear}
\end{figure}

We construct polarization bases aligned with four representative curvilinear geometries: elliptical, parabolic, bipolar, and dipole coordinate systems, Fig.~\ref{fig:curvilinear}. 
For each case, we specify the coordinate transformation, identify the corresponding conformal map, and derive the spatially varying phase required in Eq.~\ref{eq:UnitVectors} to express the basis in the circular polarization representation.

\subsection{Elliptical Coordinate System}
We define the elliptical coordinate system by the transformation,
\begin{align}
    \begin{aligned}
        x &= a \cosh u \cos v, \\
        y &= a \sinh u \sin v,
    \end{aligned}
\end{align}
where $a > 0$ places the two foci at $(x, y) = (\pm a, 0)$. 
The associated conformal map,
\begin{align}
    z = a \cosh w,
\end{align}
is one-to-one with $w = u + i v$ and domain $u \in [0, \infty)$, $v \in [0, \pi]$. 
In this geometry, curves of constant $u$ are confocal ellipses and curves of constant $v$ are confocal hyperbolas, Fig.~\ref{fig:curvilinear}(a). 
A branch cut appears along the degenerate ellipse at $u = 0$, corresponding to the segment between the foci on the $x$-axis. 
The metric becomes singular at the foci, where the scale factors vanish.

We express the basis in Eq.~(\ref{eq:UnitVectors}) using the spatially varying phase,
\begin{align}
    \theta_{u} = \arctan \left( \coth u \, \tan v \right),
\end{align}
which encodes the local geometry of the elliptical coordinates. 
This phase determines linear polarization orientations aligned with the confocal ellipses and hyperbolas.

\subsection{Parabolic Coordinate System}
We define the parabolic coordinate system by the transformation,
\begin{align}
    \begin{aligned}
        x &= \frac{1}{2} \left( u^2 - v^2 \right), \\
        y &= u v.
    \end{aligned}
\end{align}
The associated conformal map,
\begin{align}
    z = \frac{1}{2} w^2,
\end{align}
is one-to-one with $w = u + i v$ and domain $u \in [0, \infty)$ and $v \in (-\infty, \infty)$.
In this geometry, curves of constant $u$ are parabolas opening to the left, and curves of constant $v$ are parabolas opening to the right, Fig.~\ref{fig:curvilinear}(b). 
No branch cut appears in the forward map, but the inverse map $w = \sqrt{2z}$ requires one. 
The conformal map is singular only at $w = 0$, corresponding to the origin in Cartesian coordinates, where the scale factors vanish.

We express the basis in Eq.~(\ref{eq:UnitVectors}) using the spatially varying phase,
\begin{align}
    \theta_{u} = \arctan \frac{v}{u},
\end{align}
which encodes the local geometry of the parabolic coordinates. This phase determines linear polarization orientations aligned with the orthogonal families of left- and right-opening parabolas.

\subsection{Bipolar Coordinate System}
We define the bipolar coordinate system by the transformation,
\begin{align}
    \begin{aligned}
        x &= a \frac{\sinh v}{\cosh v - \cos u}, \\
        y &= a \frac{\sin u}{\cosh v - \cos u},
    \end{aligned}
\end{align}
where $a > 0$ places the two foci at $(x, y) = (\pm a, 0)$. 
The corresponding conformal map,
\begin{align}
    z = i a \cot \frac{w}{2},
\end{align}
is one-to-one with $w = u + i v$ and domain $u \in [0, 2\pi)$, $v \in (-\infty, \infty)$.
In this geometry, curves of constant $u$ are non-concentric circles centered at $(x, y) = (0, a \cot u)$ with radii $a / \sin u$, and curves of constant $v$ are non-intersecting circles centered at $(x, y) = (a \coth v, 0)$ with radii $a / \sinh v$, Fig.~\ref{fig:curvilinear}(c). 
These families of circles intersect orthogonally at the two foci.
A branch cut appears along the line segment between the foci on the $x$-axis, corresponding to $u = 0$. 
The metric becomes singular at each focal point, where the scale factors diverge.

We express the basis in Eq.~(\ref{eq:UnitVectors}) using the spatially varying phase,
\begin{align}
    \theta_{u} = \arctan \left( \frac{\cosh v \, \cos u - 1}{\sinh v \, \sin u} \right),
\end{align}
which encodes the local geometry of the bipolar coordinates. 
This phase determines linear polarization orientations aligned with the intersecting and non-intersecting families of orthogonal circles.

\subsection{Dipole Coordinate System}
We define the dipole coordinate system by the transformation,
\begin{align}
    \begin{aligned}
        x &= \frac{u}{u^{2} + v^{2}}, \\
        y &= -\frac{v}{u^{2} + v^{2}},
    \end{aligned}
\end{align}
The corresponding conformal map,
\begin{align}
    z = \frac{1}{w},
\end{align}
is one-to-one with $w = u + i v$ and domain $u \in (-\infty, 0) \cup (0, \infty)$, $v \in (-\infty, 0) \cup (0, \infty)$.
In this geometry, curves of constant $u$ are non-concentric circles centered at $(x, y) = (u/2, 0)$, and curves of constant $v$ are non-concentric circles centered at $(x, y) = (0, -v/2)$. These two families intersect orthogonally, Fig.~\ref{fig:curvilinear}(d).
No branch cut appears in the conformal map. 
However, the map is singular at $w = 0$, corresponding to the origin in Cartesian coordinates, where the scale factors diverge.

We express the basis in Eq.~(\ref{eq:UnitVectors}) using the spatially varying phase,
\begin{align}
    \theta_{u} = \arctan\left(\frac{2 u v}{v^{2} - u^{2}}\right),
\end{align}
which encodes the local geometry of the dipole coordinates. 
This phase determines linear polarization orientations aligned with the two orthogonal families of non-concentric circles.

We realize that conformal maps used to construct curvilinear polarization bases can be grouped into two broad families.

In the parabolic and dipole coordinate systems, we use algebraic maps,
\begin{align}
    z = a w^{n},
\end{align}
with $a = 1$, $n = 2$ for the parabolic system and $a = 1$, $n = -1$ for the dipole system. These power-law relations between the complex variables are known as \emph{power maps} or \emph{monomial conformal maps}.

In the elliptical and bipolar systems, we use transcendental maps,
\begin{align}
    \begin{aligned}
        z &= a \cosh b w, \\
        z &= a \coth b w,
    \end{aligned}
\end{align}
with $a > 0$, $b = 1$ for the elliptical system and $a < 0$, $b = i/2$ for the bipolar system. These maps are known as \emph{transcendental conformal maps}.
The exponential map,
\begin{align}
    z = e^{w},
\end{align}
with $u = \ln r$ for the radial coordinate, provides another example of a transcendental conformal map that generates polar coordinates.

Our framework applies to arbitrary algebraic or transcendental conformal maps and provides a systematic method to construct a wide range of analytically tractable and experimentally feasible curvilinear polarization bases.

\begin{figure}[ht]
    \centering
    \includegraphics[width= 0.75 \columnwidth]{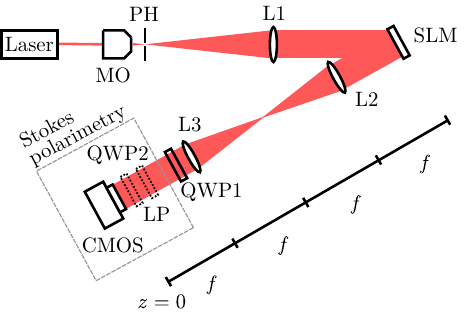}
    \caption{Experimental setup.  Laser, HeNe laser source; MO, microscope objective; PH, pinhole, L1--L3, lenses; SLM, Spatial Light Modulator; QWP1--QWP2, quarter-wave plate; LP, linear polarizer; CMOS, sensor.  The dashed box outlines the Stokes polarimetry module used for polarization state reconstruction.}
    \label{fig:setup}
\end{figure}

\section{Experimental Setup} \label{sec:Sec4}

We implemented the experimental setup shown in Fig.~\ref{fig:setup}~\cite{Otte2016} to generate the curvilinear polarization structures described above. 
The source was a diagonally polarized HeNe laser beam. 
After expansion and collimation -resulting in a beam with 2 mm radius-, the beam uniformly illuminated a phase-only SLM (Holoeye Pluto VIS, calibrated at $633~\mathrm{nm}$, with an $8~\mu\mathrm{m}$ pixel pitch). 
We configured the SLM to modulate only the horizontal component of the impinging field such that its action, expressed in the linear basis, takes the form of the Jones matrix,
\begin{align}
    \mathbf{J}_{\text{SLM}} = \begin{bmatrix}
        e^{i\theta_h(\mathbf{r})} & 0 \\
        0 & 1
    \end{bmatrix},
\end{align}
where $\theta_h(\mathbf{r})$ is the spatially varying phase applied to the horizontal polarization component by the encoded hologram, while the vertical component remains unaffected~\cite{Goodman2017,RosalesGuzman2024}.

A quarter-wave plate (QWP1) placed after the SLM, with its fast axis oriented at $45^\circ$ to the horizontal, converted the field into left- and right-circular components. 
We encoded the curvilinear polarization basis states by setting $\theta_{h}(\mathbf{r}) = 2\theta_{u}(\mathbf{r})$, as described in Sec.~\ref{sec:Sec3}. 
Lenses L2 and L3 formed a 4f relay system, imaging the SLM plane ($z = 0$) onto the detection plane.

We carried out polarization measurements using a Stokes polarimetry module~\cite{Goldstein2011}. A second quarter-wave plate (QWP2) and a linear polarizer (LP) analyzed the output field. 
Using a CMOS sensor (Nikon D7500), we acquired six intensity images corresponding to horizontal (H), vertical (V), diagonal (D), antidiagonal (A), right-circular (R), and left-circular (L) projections. 
The spatially resolved Stokes parameters were computed from these images~\cite{Cox2023},
\begin{align}
S_{0}(\mathbf{r}) &= I_{\mathrm{R}}(\mathbf{r}) + I_{\mathrm{L}}(\mathbf{r}),\\
S_{1}(\mathbf{r}) &= I_{\mathrm{H}}(\mathbf{r}) - I_{\mathrm{V}}(\mathbf{r}),\\
S_{2}(\mathbf{r}) &= I_{\mathrm{D}}(\mathbf{r}) - I_{\mathrm{A}}(\mathbf{r}),\\
S_{3}(\mathbf{r}) &= I_{\mathrm{R}}(\mathbf{r}) - I_{\mathrm{L}}(\mathbf{r}),
\end{align}
where $I_{\chi}(\mathbf{r})$ denotes the measured intensity for each polarization component $\chi \in \{ \mathrm{H}, \mathrm{V}, \mathrm{D}, \mathrm{A}, \mathrm{R}, \mathrm{L} \}$.

\section{Experimental Results} \label{sec:Sec5}

We reconstructed the curvilinear polarization bases described in Sec.~\ref{sec:Sec3} using Stokes polarimetry and compared the results with the corresponding theoretical predictions. 
Each subsection presents the experimental results for one coordinate system, following the same order as in Sec.~\ref{sec:Sec3} to facilitate direct comparison.

\begin{figure}[htp!]
    \centering
    \includegraphics[width= 0.75 \columnwidth]{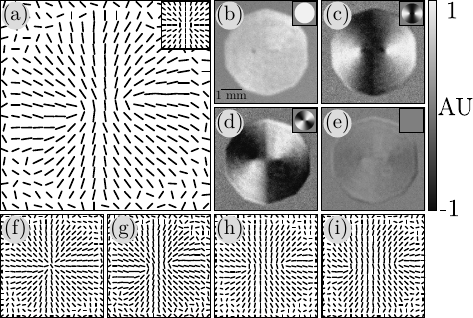}
    \caption{Experimental results for the elliptical coordinate system. 
    (a) Reconstructed polarization distribution aligned with hyperbolic coordinate curves for the semi-focal  distance $a=0.5 ~ \mathrm{mm}$.
    (b)–(e) Corresponding spatial distributions of the Stokes parameters $S_0$, $S_1$, $S_2$, and $S_3$.
    (f)–(i) Reconstructed polarization distributions for $a \in \left\{ 0.0, 0.5, 0.75, 1.0 \right\} ~ \mathrm{mm}$, in that order.
    Insets show the corresponding theoretical predictions.}
    \label{fig:elliptic}
\end{figure}

\subsection{Elliptical Coordinate System}
In the elliptical coordinate system, the semi-focal distance $a$ controls the degree of polarization modulation across the transverse plane. 
When $a$ is large, the polarization approaches a uniform state, and the elliptical pattern is suppressed. 
As $a$ decreases, the curvilinear structure emerges, and the polarization follows elliptical and hyperbolic coordinate curves.

In Fig.~\ref{fig:elliptic}, we show representative experimental results for different values of the semi-focal distance $a$.
Panel (a) shows the reconstructed polarization distribution aligned with hyperbolic coordinate curves for $a=0.5 ~\mathrm{mm}$. 
Panels (b)–(e) show the corresponding spatial distributions of the Stokes parameters $S_0$, $S_1$, $S_2$, and $S_3$, with theoretical predictions shown as insets for comparison.
Panels (f)–(i) show the reconstructed polarization distributions for increasing values of $a \in \left\{0.0, 0.5, 0.75, 1.0 \right\} ~\mathrm{mm}$, in that order, highlighting how the polarization structure evolves with focal separation.

\subsection{Parabolic Coordinate System}
\begin{figure}[ht]
    \centering
    \includegraphics[width= 0.75 \columnwidth]{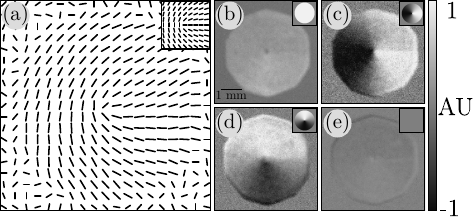}
    \caption{Experimental results for the parabolic coordinate system. 
    (a) Reconstructed polarization distribution aligned with coordinate curves corresponding to parabolas opening to the right. 
    (b)–(e) Spatial distributions of the Stokes parameters $S_0$, $S_1$, $S_2$, and $S_3$.    
    Insets show the corresponding theoretical predictions.}
    \label{fig:parabolic}
\end{figure}

In the parabolic coordinate system, we observe polarization aligned with the orthogonal families of parabolic coordinate curves opening to the right. 
Figure~\ref{fig:parabolic}(a) shows the reconstructed polarization distribution. 
Panels (b)–(e) display the spatial distributions of the Stokes parameters $S_0$, $S_1$, $S_2$, and $S_3$, with theoretical predictions shown as insets for comparison.

\subsection{Bipolar Coordinate System}
In the bipolar coordinate system, the semi-focal distance $a$ controls the degree of polarization modulation across the transverse plane. 
When $a$ is large, the polarization approaches a uniform state, and the curvilinear structure is suppressed. 
As $a$ decreases, the curvilinear structure emerges, and the polarization pattern follows the orthogonal families of non-concentric circular arcs defined by bipolar coordinates.

\begin{figure}[ht]
    \centering
    \includegraphics[width= 0.75 \columnwidth]{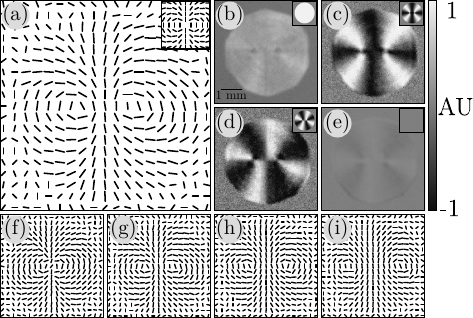}
    \caption{Experimental results for the bipolar coordinate system. 
    (a) Reconstructed polarization distribution aligned with coordinate curves corresponding to non-concentric circles for semi-focal distance $a=0.5 ~\mathrm{mm}$.
    (b)–(e) Corresponding spatial distributions of the Stokes parameters $S_0$, $S_1$, $S_2$, and $S_3$.
    (f)–(i) Reconstructed polarization distributions for $a \in \left\{0.0, 0.5, 0.75, 1.0 \right\} ~ \mathrm{mm}$, in that order.
    Insets show the corresponding theoretical predictions.}
    \label{fig:bipolar}
\end{figure}

In Fig.~\ref{fig:bipolar}, we show representative experimental results for different values of the semi-focal distance $a$.
Panel (a) shows the reconstructed polarization distribution aligned with bipolar coordinate curves  $a=0.5 ~\mathrm{mm}$.  
Panels (b)–(e) display the  corresponding spatial distributions of the Stokes parameters $S_0$, $S_1$, $S_2$, and $S_3$, with theoretical predictions shown as insets for comparison.
Panels (f)–(i) show how the reconstructed polarization distribution evolves with increasing values of $a \in \left\{0.0, 0.5, 0.75, 1.0 \right\} ~\mathrm{mm}$, in that order.

\subsection{Dipole Coordinate System}
In the dipole coordinate system, the polarization follows the orthogonal families of non-concentric circles defined by the conformal map $z = 1/w$. 
These curves concentrate angular variation near the origin, resulting in high spatial gradients in the polarization distribution.

\begin{figure}[ht]
    \centering
    \includegraphics[width= 0.75 \columnwidth]{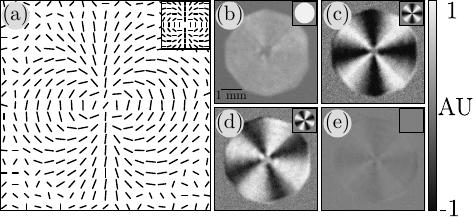}
    \caption{Experimental results for the dipole coordinate system. 
    (a) Reconstructed polarization distribution aligned with dipole coordinate curves. 
    (b)–(e) Spatial distributions of the Stokes parameters $S_0$, $S_1$, $S_2$, and $S_3$.    Insets show the corresponding theoretical predictions.}
    \label{fig:dipolar}
\end{figure}

Figure~\ref{fig:dipolar}(a) shows the reconstructed polarization distribution. 
Panels (b)–(e) display the spatial distributions of the Stokes parameters $S_0$, $S_1$, $S_2$, and $S_3$, with theoretical predictions shown as insets for comparison.

We quantified the spatial inhomogeneity of our curvilinear polarization bases using the Vector Quality Factor (VQF)~\cite{Ndagano2016},
\begin{align}
    \mathrm{VQF}_\Omega = \sqrt{1 - p_\Omega^2},
\end{align}
where $p_{\Omega}$ is the degree of polarization over a region, Eq.~\ref{eq:degree}. 
The VQF vanishes for beams with uniform polarization across the transverse plane and reaches unity for beams with spatially varying polarization distributions that average to zero net polarization.

\begin{figure}[ht]
    \centering
    \includegraphics[width= 0.75 \columnwidth]{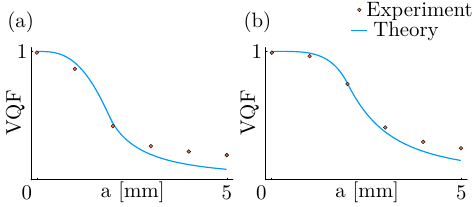}
    \caption{Vector Quality Factor (VQF) as a function of the semi-focal separation $a$ for (a) the elliptical and (b) the bipolar coordinate systems. 
    Experimental measurements and theoretical predictions are shown for both cases, the former for values $a \in \left\{ 0,1,2,3,4,5\right\} \; \mathrm{mm}$.}
    \label{fig:vqf}
\end{figure}

In the elliptical coordinate system, we evaluated the VQF as a function of the semi-focal separation $a$. 
Figure~\ref{fig:vqf}(a) shows our experimental measurements alongside theoretical predictions. We observe a smooth decay in the VQF with increasing $a$. 
At small $a$, the polarization follows hyperbolic coordinate curves, yielding $\mathrm{VQF} \approx 1$. 
As $a$ increases, the polarization becomes nearly uniform, and the VQF approaches zero.
For the parabolic coordinate system, we measured $\mathrm{VQF} = 0.99$, consistent with the theoretical value of $1$.
In the bipolar coordinate system, we evaluated the VQF as a function of $a$. 
Figure~\ref{fig:vqf}(b) shows both experimental and theoretical results.
The VQF decreases smoothly with increasing $a$, from $1$ when the polarization follows non-concentric circular arcs to near-zero for large $a$, where the distribution becomes uniform.
For the dipole coordinate system, we again measured $\mathrm{VQF} = 0.99$, in agreement with the theoretical prediction of one.

Our experimental results demonstrate that we can generate curvilinear polarization bases using conformal maps implemented through phase-only modulation. 
While we focused on selected representatives from both algebraic and transcendental families, our method applies to arbitrary conformal maps.

\section{Conclusion} \label{sec:Sec6}

We introduced a general method for constructing curvilinear polarization bases by exploiting the geometry of coordinate systems defined through conformal maps. 
Our approach enables direct encoding of spatially varying polarization structures using phase-only spatial light modulators, without relying on predefined spatial mode families.
We demonstrated the versatility of our framework through four representative examples: elliptical, parabolic, bipolar, and dipole coordinates.

We realized these polarization bases experimentally by combining phase-only modulation with Stokes polarimetry. 
The reconstructed polarization distributions match theoretical predictions across all four cases. 
These results demonstrate that our framework enables precise spatial control over polarization without constraining the field to a fixed modal basis.

The four examples span both algebraic and transcendental families of conformal maps, establishing the generality of our method. 
In two cases, their geometry enables continuous control over the spatial variation of polarization across the transverse plane. 
By connecting polarization structure directly to the geometry of curvilinear coordinates, our approach extends to arbitrary conformal mappings and provides a systematic path for generating analytically tractable and experimentally accessible vector fields. 
This connection replaces the conventional approach to vector beams using spatial modal expansions with a geometry-driven framework for polarization design.


\section*{Funding}
Not Applicable. 

\section*{Acknowledgments}
B.~M.~R.~L. acknowledges support and hospitality as an affiliate visiting colleague at the Department of Physics and Astronomy, University of New Mexico.

\section*{Disclosures}
The authors declare no conflicts of interest.

\section*{Data Availability Statement}
All the data is available from the corresponding author upon reasonable request.



\begin{thebibliography}{39}%
\makeatletter
\providecommand \@ifxundefined [1]{%
 \@ifx{#1\undefined}
}%
\providecommand \@ifnum [1]{%
 \ifnum #1\expandafter \@firstoftwo
 \else \expandafter \@secondoftwo
 \fi
}%
\providecommand \@ifx [1]{%
 \ifx #1\expandafter \@firstoftwo
 \else \expandafter \@secondoftwo
 \fi
}%
\providecommand \natexlab [1]{#1}%
\providecommand \enquote  [1]{``#1''}%
\providecommand \bibnamefont  [1]{#1}%
\providecommand \bibfnamefont [1]{#1}%
\providecommand \citenamefont [1]{#1}%
\providecommand \href@noop [0]{\@secondoftwo}%
\providecommand \href [0]{\begingroup \@sanitize@url \@href}%
\providecommand \@href[1]{\@@startlink{#1}\@@href}%
\providecommand \@@href[1]{\endgroup#1\@@endlink}%
\providecommand \@sanitize@url [0]{\catcode `\\12\catcode `\$12\catcode `\&12\catcode `\#12\catcode `\^12\catcode `\_12\catcode `\%12\relax}%
\providecommand \@@startlink[1]{}%
\providecommand \@@endlink[0]{}%
\providecommand \url  [0]{\begingroup\@sanitize@url \@url }%
\providecommand \@url [1]{\endgroup\@href {#1}{\urlprefix }}%
\providecommand \urlprefix  [0]{URL }%
\providecommand \Eprint [0]{\href }%
\providecommand \doibase [0]{https://doi.org/}%
\providecommand \selectlanguage [0]{\@gobble}%
\providecommand \bibinfo  [0]{\@secondoftwo}%
\providecommand \bibfield  [0]{\@secondoftwo}%
\providecommand \translation [1]{[#1]}%
\providecommand \BibitemOpen [0]{}%
\providecommand \bibitemStop [0]{}%
\providecommand \bibitemNoStop [0]{.\EOS\space}%
\providecommand \EOS [0]{\spacefactor3000\relax}%
\providecommand \BibitemShut  [1]{\csname bibitem#1\endcsname}%
\let\auto@bib@innerbib\@empty
\bibitem [{\citenamefont {Rubinsztein-Dunlop}\ \emph {et~al.}(2017)\citenamefont {Rubinsztein-Dunlop}, \citenamefont {Forbes}, \citenamefont {Berry}, \citenamefont {Dennis}, \citenamefont {Andrews}, \citenamefont {Mansuripur}, \citenamefont {Denz}, \citenamefont {Alpmann}, \citenamefont {Banzer}, \citenamefont {Bauer}, \citenamefont {Karimi}, \citenamefont {Marrucci}, \citenamefont {Padgett}, \citenamefont {Ritsch-Marte}, \citenamefont {Litchinitser}, \citenamefont {Bigelow}, \citenamefont {Rosales-Guzm{\'a}n}, \citenamefont {Belmonte}, \citenamefont {Torres}, \citenamefont {Neely}, \citenamefont {Baker}, \citenamefont {Gordon}, \citenamefont {Stilgoe}, \citenamefont {Romero}, \citenamefont {White}, \citenamefont {Fickler}, \citenamefont {Willner}, \citenamefont {Xie}, \citenamefont {McMorran},\ and\ \citenamefont {Weiner}}]{Roadmap2017}%
  \BibitemOpen
  \bibfield  {author} {\bibinfo {author} {\bibfnamefont {H.}~\bibnamefont {Rubinsztein-Dunlop}}, \bibinfo {author} {\bibfnamefont {A.}~\bibnamefont {Forbes}}, \bibinfo {author} {\bibfnamefont {M.~V.}\ \bibnamefont {Berry}}, \bibinfo {author} {\bibfnamefont {M.~R.}\ \bibnamefont {Dennis}}, \bibinfo {author} {\bibfnamefont {D.~L.}\ \bibnamefont {Andrews}}, \bibinfo {author} {\bibfnamefont {M.}~\bibnamefont {Mansuripur}}, \bibinfo {author} {\bibfnamefont {C.}~\bibnamefont {Denz}}, \bibinfo {author} {\bibfnamefont {C.}~\bibnamefont {Alpmann}}, \bibinfo {author} {\bibfnamefont {P.}~\bibnamefont {Banzer}}, \bibinfo {author} {\bibfnamefont {T.}~\bibnamefont {Bauer}}, \bibinfo {author} {\bibfnamefont {E.}~\bibnamefont {Karimi}}, \bibinfo {author} {\bibfnamefont {L.}~\bibnamefont {Marrucci}}, \bibinfo {author} {\bibfnamefont {M.}~\bibnamefont {Padgett}}, \bibinfo {author} {\bibfnamefont {M.}~\bibnamefont {Ritsch-Marte}}, \bibinfo {author} {\bibfnamefont {N.~M.}\ \bibnamefont {Litchinitser}}, \bibinfo {author}
  {\bibfnamefont {N.~P.}\ \bibnamefont {Bigelow}}, \bibinfo {author} {\bibfnamefont {C.}~\bibnamefont {Rosales-Guzm{\'a}n}}, \bibinfo {author} {\bibfnamefont {A.}~\bibnamefont {Belmonte}}, \bibinfo {author} {\bibfnamefont {J.~P.}\ \bibnamefont {Torres}}, \bibinfo {author} {\bibfnamefont {T.~W.}\ \bibnamefont {Neely}}, \bibinfo {author} {\bibfnamefont {M.}~\bibnamefont {Baker}}, \bibinfo {author} {\bibfnamefont {R.}~\bibnamefont {Gordon}}, \bibinfo {author} {\bibfnamefont {A.~B.}\ \bibnamefont {Stilgoe}}, \bibinfo {author} {\bibfnamefont {J.}~\bibnamefont {Romero}}, \bibinfo {author} {\bibfnamefont {A.~G.}\ \bibnamefont {White}}, \bibinfo {author} {\bibfnamefont {R.}~\bibnamefont {Fickler}}, \bibinfo {author} {\bibfnamefont {A.~E.}\ \bibnamefont {Willner}}, \bibinfo {author} {\bibfnamefont {G.}~\bibnamefont {Xie}}, \bibinfo {author} {\bibfnamefont {B.}~\bibnamefont {McMorran}},\ and\ \bibinfo {author} {\bibfnamefont {A.~M.}\ \bibnamefont {Weiner}},\ }\bibfield  {title} {\bibinfo {title} {Roadmap on structured
  light},\ }\href {https://doi.org/10.1088/2040-8978/19/1/013001} {\bibfield  {journal} {\bibinfo  {journal} {J. Opt.}\ }\textbf {\bibinfo {volume} {19}},\ \bibinfo {pages} {013001} (\bibinfo {year} {2017})}\BibitemShut {NoStop}%
\bibitem [{\citenamefont {Rosales-Guzm{\'a}n}\ and\ \citenamefont {Rodr{\'\i}guez-Fajardo}(2024)}]{RosalesGuzman2024b}%
  \BibitemOpen
  \bibfield  {author} {\bibinfo {author} {\bibfnamefont {C.}~\bibnamefont {Rosales-Guzm{\'a}n}}\ and\ \bibinfo {author} {\bibfnamefont {V.}~\bibnamefont {Rodr{\'\i}guez-Fajardo}},\ }\bibfield  {title} {\bibinfo {title} {A perspective on structured light's applications},\ }\href {https://doi.org/10.1063/5.0236477} {\bibfield  {journal} {\bibinfo  {journal} {Appl. Phys. Lett.}\ }\textbf {\bibinfo {volume} {125}},\ \bibinfo {pages} {200503} (\bibinfo {year} {2024})}\BibitemShut {NoStop}%
\bibitem [{\citenamefont {Forbes}\ \emph {et~al.}(2019)\citenamefont {Forbes}, \citenamefont {Aiello},\ and\ \citenamefont {Ndagano}}]{Forbes2019}%
  \BibitemOpen
  \bibfield  {author} {\bibinfo {author} {\bibfnamefont {A.}~\bibnamefont {Forbes}}, \bibinfo {author} {\bibfnamefont {A.}~\bibnamefont {Aiello}},\ and\ \bibinfo {author} {\bibfnamefont {B.}~\bibnamefont {Ndagano}},\ }\bibfield  {title} {\bibinfo {title} {Classically entangled light},\ }in\ \href {https://doi.org/10.1016/bs.po.2018.11.001} {\emph {\bibinfo {booktitle} {Progress in Optics}}},\ Vol.~\bibinfo {volume} {64},\ \bibinfo {editor} {edited by\ \bibinfo {editor} {\bibfnamefont {T.~D.}\ \bibnamefont {Visser}}}\ (\bibinfo  {publisher} {Elsevier},\ \bibinfo {address} {Amsterdam},\ \bibinfo {year} {2019})\ pp.\ \bibinfo {pages} {99--153}\BibitemShut {NoStop}%
\bibitem [{\citenamefont {Shen}\ \emph {et~al.}(2024)\citenamefont {Shen}, \citenamefont {Zhang}, \citenamefont {Shi}, \citenamefont {Du}, \citenamefont {Yuan},\ and\ \citenamefont {Zayats}}]{Shen2024}%
  \BibitemOpen
  \bibfield  {author} {\bibinfo {author} {\bibfnamefont {Y.}~\bibnamefont {Shen}}, \bibinfo {author} {\bibfnamefont {Q.}~\bibnamefont {Zhang}}, \bibinfo {author} {\bibfnamefont {P.}~\bibnamefont {Shi}}, \bibinfo {author} {\bibfnamefont {L.}~\bibnamefont {Du}}, \bibinfo {author} {\bibfnamefont {X.}~\bibnamefont {Yuan}},\ and\ \bibinfo {author} {\bibfnamefont {A.~V.}\ \bibnamefont {Zayats}},\ }\bibfield  {title} {\bibinfo {title} {Optical skyrmions and other topological quasiparticles of light},\ }\href {https://doi.org/10.1038/s41566-023-01325-7} {\bibfield  {journal} {\bibinfo  {journal} {Nat. Photonics}\ }\textbf {\bibinfo {volume} {18}},\ \bibinfo {pages} {15} (\bibinfo {year} {2024})}\BibitemShut {NoStop}%
\bibitem [{\citenamefont {Rosales-Guzm{\'a}n}\ \emph {et~al.}(2018)\citenamefont {Rosales-Guzm{\'a}n}, \citenamefont {Ndagano},\ and\ \citenamefont {Forbes}}]{RosalesGuzman2018}%
  \BibitemOpen
  \bibfield  {author} {\bibinfo {author} {\bibfnamefont {C.}~\bibnamefont {Rosales-Guzm{\'a}n}}, \bibinfo {author} {\bibfnamefont {B.}~\bibnamefont {Ndagano}},\ and\ \bibinfo {author} {\bibfnamefont {A.}~\bibnamefont {Forbes}},\ }\bibfield  {title} {\bibinfo {title} {A review of complex vector light fields and their applications},\ }\href {https://doi.org/10.1088/2040-8986/aaeb7d} {\bibfield  {journal} {\bibinfo  {journal} {J. Opt.}\ }\textbf {\bibinfo {volume} {20}},\ \bibinfo {pages} {123001} (\bibinfo {year} {2018})}\BibitemShut {NoStop}%
\bibitem [{\citenamefont {Shen}\ and\ \citenamefont {Rosales-Guzm{\'a}n}(2022)}]{Shen2022}%
  \BibitemOpen
  \bibfield  {author} {\bibinfo {author} {\bibfnamefont {Y.}~\bibnamefont {Shen}}\ and\ \bibinfo {author} {\bibfnamefont {C.}~\bibnamefont {Rosales-Guzm{\'a}n}},\ }\bibfield  {title} {\bibinfo {title} {Nonseparable states of light: from quantum to classical},\ }\href {https://doi.org/10.1002/lpor.202100533} {\bibfield  {journal} {\bibinfo  {journal} {Laser Photonics Rev.}\ }\textbf {\bibinfo {volume} {16}},\ \bibinfo {pages} {2100533} (\bibinfo {year} {2022})}\BibitemShut {NoStop}%
\bibitem [{\citenamefont {Rosales-Guzm{\'a}n}\ and\ \citenamefont {Forbes}(2024)}]{RosalesGuzman2024}%
  \BibitemOpen
  \bibfield  {author} {\bibinfo {author} {\bibfnamefont {C.}~\bibnamefont {Rosales-Guzm{\'a}n}}\ and\ \bibinfo {author} {\bibfnamefont {A.}~\bibnamefont {Forbes}},\ }\href {https://doi.org/10.1117/3.100024} {\emph {\bibinfo {title} {Structured Light with Spatial Light Modulators}}},\ \bibinfo {edition} {2nd}\ ed.\ (\bibinfo  {publisher} {SPIE},\ \bibinfo {address} {Bellingham},\ \bibinfo {year} {2024})\BibitemShut {NoStop}%
\bibitem [{\citenamefont {Hu}\ and\ \citenamefont {Rosales-Guzm{\'a}n}(2022)}]{Hu2022}%
  \BibitemOpen
  \bibfield  {author} {\bibinfo {author} {\bibfnamefont {X.~B.}\ \bibnamefont {Hu}}\ and\ \bibinfo {author} {\bibfnamefont {C.}~\bibnamefont {Rosales-Guzm{\'a}n}},\ }\bibfield  {title} {\bibinfo {title} {Generation and characterization of complex vector modes with digital micromirror devices: a tutorial},\ }\href {https://doi.org/10.1088/2040-8986/ac4f0b} {\bibfield  {journal} {\bibinfo  {journal} {J. Opt.}\ }\textbf {\bibinfo {volume} {24}},\ \bibinfo {pages} {034001} (\bibinfo {year} {2022})}\BibitemShut {NoStop}%
\bibitem [{\citenamefont {Scholes}\ \emph {et~al.}(2019)\citenamefont {Scholes}, \citenamefont {Kara}, \citenamefont {Pinnell}, \citenamefont {Rodr{\'\i}guez-Fajardo},\ and\ \citenamefont {Forbes}}]{Scholes2019}%
  \BibitemOpen
  \bibfield  {author} {\bibinfo {author} {\bibfnamefont {S.}~\bibnamefont {Scholes}}, \bibinfo {author} {\bibfnamefont {R.}~\bibnamefont {Kara}}, \bibinfo {author} {\bibfnamefont {J.}~\bibnamefont {Pinnell}}, \bibinfo {author} {\bibfnamefont {V.}~\bibnamefont {Rodr{\'\i}guez-Fajardo}},\ and\ \bibinfo {author} {\bibfnamefont {A.}~\bibnamefont {Forbes}},\ }\bibfield  {title} {\bibinfo {title} {Structured light with digital micromirror devices: a guide to best practice},\ }\href {https://doi.org/10.1117/1.OE.59.4.041202} {\bibfield  {journal} {\bibinfo  {journal} {Opt. Eng.}\ }\textbf {\bibinfo {volume} {59}},\ \bibinfo {pages} {041202} (\bibinfo {year} {2019})}\BibitemShut {NoStop}%
\bibitem [{\citenamefont {Maurer}\ \emph {et~al.}(2007)\citenamefont {Maurer}, \citenamefont {Jesacher}, \citenamefont {F{\"u}rhapter}, \citenamefont {Bernet},\ and\ \citenamefont {Ritsch-Marte}}]{Maurer2007}%
  \BibitemOpen
  \bibfield  {author} {\bibinfo {author} {\bibfnamefont {C.}~\bibnamefont {Maurer}}, \bibinfo {author} {\bibfnamefont {A.}~\bibnamefont {Jesacher}}, \bibinfo {author} {\bibfnamefont {S.}~\bibnamefont {F{\"u}rhapter}}, \bibinfo {author} {\bibfnamefont {S.}~\bibnamefont {Bernet}},\ and\ \bibinfo {author} {\bibfnamefont {M.}~\bibnamefont {Ritsch-Marte}},\ }\bibfield  {title} {\bibinfo {title} {Tailoring of arbitrary optical vector beams},\ }\href {https://doi.org/10.1088/1367-2630/9/3/078} {\bibfield  {journal} {\bibinfo  {journal} {New J. Phys.}\ }\textbf {\bibinfo {volume} {9}},\ \bibinfo {pages} {78} (\bibinfo {year} {2007})}\BibitemShut {NoStop}%
\bibitem [{\citenamefont {Niziev}\ \emph {et~al.}(2006)\citenamefont {Niziev}, \citenamefont {Chang},\ and\ \citenamefont {Nesterov}}]{Niziev2006}%
  \BibitemOpen
  \bibfield  {author} {\bibinfo {author} {\bibfnamefont {V.~G.}\ \bibnamefont {Niziev}}, \bibinfo {author} {\bibfnamefont {R.~S.}\ \bibnamefont {Chang}},\ and\ \bibinfo {author} {\bibfnamefont {A.~V.}\ \bibnamefont {Nesterov}},\ }\bibfield  {title} {\bibinfo {title} {Generation of inhomogeneously polarized laser beams by use of a {Sagnac} interferometer},\ }\href {https://doi.org/10.1364/AO.45.008393} {\bibfield  {journal} {\bibinfo  {journal} {Appl. Opt.}\ }\textbf {\bibinfo {volume} {45}},\ \bibinfo {pages} {8393} (\bibinfo {year} {2006})}\BibitemShut {NoStop}%
\bibitem [{\citenamefont {Rong}\ \emph {et~al.}(2014)\citenamefont {Rong}, \citenamefont {Han}, \citenamefont {Wang},\ and\ \citenamefont {Guo}}]{Rong2014}%
  \BibitemOpen
  \bibfield  {author} {\bibinfo {author} {\bibfnamefont {Z.~Y.}\ \bibnamefont {Rong}}, \bibinfo {author} {\bibfnamefont {Y.~J.}\ \bibnamefont {Han}}, \bibinfo {author} {\bibfnamefont {S.~Z.}\ \bibnamefont {Wang}},\ and\ \bibinfo {author} {\bibfnamefont {C.}~\bibnamefont {Guo}},\ }\bibfield  {title} {\bibinfo {title} {Generation of arbitrary vector beams with cascaded liquid crystal spatial light modulators},\ }\href {https://doi.org/10.1364/OE.22.001636} {\bibfield  {journal} {\bibinfo  {journal} {Opt. Express}\ }\textbf {\bibinfo {volume} {22}},\ \bibinfo {pages} {1636} (\bibinfo {year} {2014})}\BibitemShut {NoStop}%
\bibitem [{\citenamefont {Otte}\ \emph {et~al.}(2018)\citenamefont {Otte}, \citenamefont {Tekce},\ and\ \citenamefont {Denz}}]{Otte2018b}%
  \BibitemOpen
  \bibfield  {author} {\bibinfo {author} {\bibfnamefont {E.}~\bibnamefont {Otte}}, \bibinfo {author} {\bibfnamefont {K.}~\bibnamefont {Tekce}},\ and\ \bibinfo {author} {\bibfnamefont {C.}~\bibnamefont {Denz}},\ }\bibfield  {title} {\bibinfo {title} {Spatial multiplexing for tailored fully-structured light},\ }\href {https://doi.org/10.1088/2040-8986/aada9b} {\bibfield  {journal} {\bibinfo  {journal} {J. Opt.}\ }\textbf {\bibinfo {volume} {20}},\ \bibinfo {pages} {105606} (\bibinfo {year} {2018})}\BibitemShut {NoStop}%
\bibitem [{\citenamefont {Rodr{\'\i}guez-Fajardo}\ \emph {et~al.}(2024{\natexlab{a}})\citenamefont {Rodr{\'\i}guez-Fajardo}, \citenamefont {Arvizu}, \citenamefont {Daza-Salgado}, \citenamefont {Perez-Garcia},\ and\ \citenamefont {Rosales-Guzm{\'a}n}}]{RodriguezFajardo2024}%
  \BibitemOpen
  \bibfield  {author} {\bibinfo {author} {\bibfnamefont {V.}~\bibnamefont {Rodr{\'\i}guez-Fajardo}}, \bibinfo {author} {\bibfnamefont {F.}~\bibnamefont {Arvizu}}, \bibinfo {author} {\bibfnamefont {D.}~\bibnamefont {Daza-Salgado}}, \bibinfo {author} {\bibfnamefont {B.}~\bibnamefont {Perez-Garcia}},\ and\ \bibinfo {author} {\bibfnamefont {C.}~\bibnamefont {Rosales-Guzm{\'a}n}},\ }\bibfield  {title} {\bibinfo {title} {On-axis complex-amplitude modulation for the generation of super-stable vector modes},\ }\href {https://doi.org/10.1088/2040-8986/ad4654} {\bibfield  {journal} {\bibinfo  {journal} {J. Opt.}\ }\textbf {\bibinfo {volume} {26}},\ \bibinfo {pages} {065606} (\bibinfo {year} {2024}{\natexlab{a}})}\BibitemShut {NoStop}%
\bibitem [{\citenamefont {Rosales-Guzm{\'a}n}\ \emph {et~al.}(2020)\citenamefont {Rosales-Guzm{\'a}n}, \citenamefont {Hu}, \citenamefont {Selyem}, \citenamefont {Moreno-Acosta}, \citenamefont {Franke-Arnold}, \citenamefont {Ramos-Garcia},\ and\ \citenamefont {Forbes}}]{Rosales2020}%
  \BibitemOpen
  \bibfield  {author} {\bibinfo {author} {\bibfnamefont {C.}~\bibnamefont {Rosales-Guzm{\'a}n}}, \bibinfo {author} {\bibfnamefont {X.~B.}\ \bibnamefont {Hu}}, \bibinfo {author} {\bibfnamefont {A.}~\bibnamefont {Selyem}}, \bibinfo {author} {\bibfnamefont {P.}~\bibnamefont {Moreno-Acosta}}, \bibinfo {author} {\bibfnamefont {S.}~\bibnamefont {Franke-Arnold}}, \bibinfo {author} {\bibfnamefont {R.}~\bibnamefont {Ramos-Garcia}},\ and\ \bibinfo {author} {\bibfnamefont {A.}~\bibnamefont {Forbes}},\ }\bibfield  {title} {\bibinfo {title} {Polarisation-insensitive generation of complex vector modes from a digital micromirror device},\ }\href {https://doi.org/10.1038/s41598-020-66799-9} {\bibfield  {journal} {\bibinfo  {journal} {Sci. Rep.}\ }\textbf {\bibinfo {volume} {10}},\ \bibinfo {pages} {10434} (\bibinfo {year} {2020})}\BibitemShut {NoStop}%
\bibitem [{\citenamefont {J{\'a}uregui}\ and\ \citenamefont {Hacyan}(2005)}]{Jauregui2005}%
  \BibitemOpen
  \bibfield  {author} {\bibinfo {author} {\bibfnamefont {R.}~\bibnamefont {J{\'a}uregui}}\ and\ \bibinfo {author} {\bibfnamefont {S.}~\bibnamefont {Hacyan}},\ }\bibfield  {title} {\bibinfo {title} {Quantum-mechanical properties of {Bessel} beams},\ }\href {https://doi.org/10.1103/PhysRevA.71.033411} {\bibfield  {journal} {\bibinfo  {journal} {Phys. Rev. A}\ }\textbf {\bibinfo {volume} {71}},\ \bibinfo {pages} {033411} (\bibinfo {year} {2005})}\BibitemShut {NoStop}%
\bibitem [{\citenamefont {Rodr\'{\i}guez-Lara}\ and\ \citenamefont {J\'auregui}(2008)}]{RodriguezLara2008}%
  \BibitemOpen
  \bibfield  {author} {\bibinfo {author} {\bibfnamefont {B.~M.}\ \bibnamefont {Rodr\'{\i}guez-Lara}}\ and\ \bibinfo {author} {\bibfnamefont {R.}~\bibnamefont {J\'auregui}},\ }\bibfield  {title} {\bibinfo {title} {Dynamical constants for electromagnetic fields with elliptic-cylindrical symmetry},\ }\href {https://doi.org/10.1103/PhysRevA.78.033813} {\bibfield  {journal} {\bibinfo  {journal} {Phys. Rev. A}\ }\textbf {\bibinfo {volume} {78}},\ \bibinfo {pages} {033813} (\bibinfo {year} {2008})}\BibitemShut {NoStop}%
\bibitem [{\citenamefont {{Rodr{\'i}guez-Lara}}\ and\ \citenamefont {J{\'a}uregui}(2009)}]{RodriguezLara2009}%
  \BibitemOpen
  \bibfield  {author} {\bibinfo {author} {\bibfnamefont {B.~M.}\ \bibnamefont {{Rodr{\'i}guez-Lara}}}\ and\ \bibinfo {author} {\bibfnamefont {R.}~\bibnamefont {J{\'a}uregui}},\ }\bibfield  {title} {\bibinfo {title} {Dynamical constants of structured photons with parabolic-cylindrical symmetry},\ }\href {https://doi.org/10.1103/PhysRevA.79.055806} {\bibfield  {journal} {\bibinfo  {journal} {Phys. Rev. A}\ }\textbf {\bibinfo {volume} {79}},\ \bibinfo {pages} {055806} (\bibinfo {year} {2009})}\BibitemShut {NoStop}%
\bibitem [{\citenamefont {Siegman}(1986)}]{Siegman1986}%
  \BibitemOpen
  \bibfield  {author} {\bibinfo {author} {\bibfnamefont {A.~E.}\ \bibnamefont {Siegman}},\ }\href@noop {} {\emph {\bibinfo {title} {Lasers}}}\ (\bibinfo  {publisher} {University Science Books},\ \bibinfo {address} {Mill Valley, CA},\ \bibinfo {year} {1986})\BibitemShut {NoStop}%
\bibitem [{\citenamefont {Bandres}\ and\ \citenamefont {Guti{\'e}rrez-Vega}(2004)}]{Bandres2004}%
  \BibitemOpen
  \bibfield  {author} {\bibinfo {author} {\bibfnamefont {M.~A.}\ \bibnamefont {Bandres}}\ and\ \bibinfo {author} {\bibfnamefont {J.~C.}\ \bibnamefont {Guti{\'e}rrez-Vega}},\ }\bibfield  {title} {\bibinfo {title} {{Ince--Gaussian} beams},\ }\href {https://doi.org/10.1364/OL.29.000144} {\bibfield  {journal} {\bibinfo  {journal} {Opt. Lett.}\ }\textbf {\bibinfo {volume} {29}},\ \bibinfo {pages} {144} (\bibinfo {year} {2004})}\BibitemShut {NoStop}%
\bibitem [{\citenamefont {Dudley}\ \emph {et~al.}(2013)\citenamefont {Dudley}, \citenamefont {Li}, \citenamefont {Mhlanga}, \citenamefont {Escuti},\ and\ \citenamefont {Forbes}}]{Dudley2013}%
  \BibitemOpen
  \bibfield  {author} {\bibinfo {author} {\bibfnamefont {A.}~\bibnamefont {Dudley}}, \bibinfo {author} {\bibfnamefont {Y.}~\bibnamefont {Li}}, \bibinfo {author} {\bibfnamefont {T.}~\bibnamefont {Mhlanga}}, \bibinfo {author} {\bibfnamefont {M.}~\bibnamefont {Escuti}},\ and\ \bibinfo {author} {\bibfnamefont {A.}~\bibnamefont {Forbes}},\ }\bibfield  {title} {\bibinfo {title} {Generating and measuring nondiffracting vector {Bessel} beams},\ }\href {https://doi.org/10.1364/OL.38.003429} {\bibfield  {journal} {\bibinfo  {journal} {Opt. Lett.}\ }\textbf {\bibinfo {volume} {38}},\ \bibinfo {pages} {3429} (\bibinfo {year} {2013})}\BibitemShut {NoStop}%
\bibitem [{\citenamefont {Galvez}\ \emph {et~al.}(2012)\citenamefont {Galvez}, \citenamefont {Khadka}, \citenamefont {Schubert},\ and\ \citenamefont {Nomoto}}]{Galvez2012}%
  \BibitemOpen
  \bibfield  {author} {\bibinfo {author} {\bibfnamefont {E.~J.}\ \bibnamefont {Galvez}}, \bibinfo {author} {\bibfnamefont {S.}~\bibnamefont {Khadka}}, \bibinfo {author} {\bibfnamefont {W.~H.}\ \bibnamefont {Schubert}},\ and\ \bibinfo {author} {\bibfnamefont {S.}~\bibnamefont {Nomoto}},\ }\bibfield  {title} {\bibinfo {title} {{Poincar{\'e}} beam patterns produced by nonseparable superpositions of {Laguerre-Gauss} and polarization modes of light},\ }\href {https://doi.org/10.1364/AO.51.002925} {\bibfield  {journal} {\bibinfo  {journal} {Appl. Opt.}\ }\textbf {\bibinfo {volume} {51}},\ \bibinfo {pages} {2925} (\bibinfo {year} {2012})}\BibitemShut {NoStop}%
\bibitem [{\citenamefont {Rosales-Guzm{\'a}n}\ \emph {et~al.}(2021)\citenamefont {Rosales-Guzm{\'a}n}, \citenamefont {Hu}, \citenamefont {Rodr{\'\i}guez-Fajardo}, \citenamefont {Hernandez-Aranda}, \citenamefont {Forbes},\ and\ \citenamefont {Perez-Garcia}}]{Rosales2021Mathieu}%
  \BibitemOpen
  \bibfield  {author} {\bibinfo {author} {\bibfnamefont {C.}~\bibnamefont {Rosales-Guzm{\'a}n}}, \bibinfo {author} {\bibfnamefont {X.~B.}\ \bibnamefont {Hu}}, \bibinfo {author} {\bibfnamefont {V.}~\bibnamefont {Rodr{\'\i}guez-Fajardo}}, \bibinfo {author} {\bibfnamefont {R.~I.}\ \bibnamefont {Hernandez-Aranda}}, \bibinfo {author} {\bibfnamefont {A.}~\bibnamefont {Forbes}},\ and\ \bibinfo {author} {\bibfnamefont {B.}~\bibnamefont {Perez-Garcia}},\ }\bibfield  {title} {\bibinfo {title} {Experimental generation of helical {Mathieu--Gauss} vector modes},\ }\href {https://doi.org/10.1088/2040-8986/abe1a8} {\bibfield  {journal} {\bibinfo  {journal} {J. Opt.}\ }\textbf {\bibinfo {volume} {23}},\ \bibinfo {pages} {034004} (\bibinfo {year} {2021})}\BibitemShut {NoStop}%
\bibitem [{\citenamefont {Li}\ \emph {et~al.}(2020)\citenamefont {Li}, \citenamefont {Hu}, \citenamefont {Perez-Garcia}, \citenamefont {Zhao}, \citenamefont {Gao}, \citenamefont {Zhu},\ and\ \citenamefont {Rosales-Guzm{\'a}n}}]{Liyao2020}%
  \BibitemOpen
  \bibfield  {author} {\bibinfo {author} {\bibfnamefont {Y.}~\bibnamefont {Li}}, \bibinfo {author} {\bibfnamefont {X.~B.}\ \bibnamefont {Hu}}, \bibinfo {author} {\bibfnamefont {B.}~\bibnamefont {Perez-Garcia}}, \bibinfo {author} {\bibfnamefont {B.}~\bibnamefont {Zhao}}, \bibinfo {author} {\bibfnamefont {W.}~\bibnamefont {Gao}}, \bibinfo {author} {\bibfnamefont {Z.~H.}\ \bibnamefont {Zhu}},\ and\ \bibinfo {author} {\bibfnamefont {C.}~\bibnamefont {Rosales-Guzm{\'a}n}},\ }\bibfield  {title} {\bibinfo {title} {Classically entangled {Ince--Gaussian} modes},\ }\href {https://doi.org/10.1063/5.0009932} {\bibfield  {journal} {\bibinfo  {journal} {Appl. Phys. Lett.}\ }\textbf {\bibinfo {volume} {116}},\ \bibinfo {pages} {221105} (\bibinfo {year} {2020})}\BibitemShut {NoStop}%
\bibitem [{\citenamefont {Zhao}\ \emph {et~al.}(2022)\citenamefont {Zhao}, \citenamefont {Rodr{\'\i}guez-Fajardo}, \citenamefont {Hu}, \citenamefont {Hernandez-Aranda}, \citenamefont {Perez-Garcia},\ and\ \citenamefont {Rosales-Guzm{\'a}n}}]{ZhaoBo2022}%
  \BibitemOpen
  \bibfield  {author} {\bibinfo {author} {\bibfnamefont {B.}~\bibnamefont {Zhao}}, \bibinfo {author} {\bibfnamefont {V.}~\bibnamefont {Rodr{\'\i}guez-Fajardo}}, \bibinfo {author} {\bibfnamefont {X.~B.}\ \bibnamefont {Hu}}, \bibinfo {author} {\bibfnamefont {R.~I.}\ \bibnamefont {Hernandez-Aranda}}, \bibinfo {author} {\bibfnamefont {B.}~\bibnamefont {Perez-Garcia}},\ and\ \bibinfo {author} {\bibfnamefont {C.}~\bibnamefont {Rosales-Guzm{\'a}n}},\ }\bibfield  {title} {\bibinfo {title} {Parabolic-accelerating vector waves},\ }\href {https://doi.org/10.1515/nanoph-2021-0695} {\bibfield  {journal} {\bibinfo  {journal} {Nanophotonics}\ }\textbf {\bibinfo {volume} {11}},\ \bibinfo {pages} {681} (\bibinfo {year} {2022})}\BibitemShut {NoStop}%
\bibitem [{\citenamefont {Medina-Segura}\ \emph {et~al.}(2023)\citenamefont {Medina-Segura}, \citenamefont {Miranda-Culin}, \citenamefont {Rodr{\'\i}guez-Fajardo}, \citenamefont {Perez-Garcia},\ and\ \citenamefont {Rosales-Guzm{\'a}n}}]{MedinaSegura2023}%
  \BibitemOpen
  \bibfield  {author} {\bibinfo {author} {\bibfnamefont {E.}~\bibnamefont {Medina-Segura}}, \bibinfo {author} {\bibfnamefont {L.}~\bibnamefont {Miranda-Culin}}, \bibinfo {author} {\bibfnamefont {V.}~\bibnamefont {Rodr{\'\i}guez-Fajardo}}, \bibinfo {author} {\bibfnamefont {B.}~\bibnamefont {Perez-Garcia}},\ and\ \bibinfo {author} {\bibfnamefont {C.}~\bibnamefont {Rosales-Guzm{\'a}n}},\ }\bibfield  {title} {\bibinfo {title} {Helico-conical vector beams},\ }\href {https://doi.org/10.1364/OL.492146} {\bibfield  {journal} {\bibinfo  {journal} {Opt. Lett.}\ }\textbf {\bibinfo {volume} {48}},\ \bibinfo {pages} {4897} (\bibinfo {year} {2023})}\BibitemShut {NoStop}%
\bibitem [{\citenamefont {Rodr{\'\i}guez-Fajardo}\ \emph {et~al.}(2024{\natexlab{b}})\citenamefont {Rodr{\'\i}guez-Fajardo}, \citenamefont {Flores-Cova}, \citenamefont {Rosales-Guzm{\'a}n},\ and\ \citenamefont {Perez-Garcia}}]{RodriguezFajardo2024Pearcey}%
  \BibitemOpen
  \bibfield  {author} {\bibinfo {author} {\bibfnamefont {V.}~\bibnamefont {Rodr{\'\i}guez-Fajardo}}, \bibinfo {author} {\bibfnamefont {G.}~\bibnamefont {Flores-Cova}}, \bibinfo {author} {\bibfnamefont {C.}~\bibnamefont {Rosales-Guzm{\'a}n}},\ and\ \bibinfo {author} {\bibfnamefont {B.}~\bibnamefont {Perez-Garcia}},\ }\bibfield  {title} {\bibinfo {title} {Experimental generation of scalar and vector vortex {Pearcey--Gauss} beams},\ }\href {https://doi.org/10.1088/2515-7647/ad452f} {\bibfield  {journal} {\bibinfo  {journal} {J. Phys. Photon.}\ }\textbf {\bibinfo {volume} {6}},\ \bibinfo {pages} {045015} (\bibinfo {year} {2024}{\natexlab{b}})}\BibitemShut {NoStop}%
\bibitem [{\citenamefont {Tidwell}\ \emph {et~al.}(1990)\citenamefont {Tidwell}, \citenamefont {Ford},\ and\ \citenamefont {Kimura}}]{Tidwell1990}%
  \BibitemOpen
  \bibfield  {author} {\bibinfo {author} {\bibfnamefont {S.~C.}\ \bibnamefont {Tidwell}}, \bibinfo {author} {\bibfnamefont {D.~H.}\ \bibnamefont {Ford}},\ and\ \bibinfo {author} {\bibfnamefont {W.~D.}\ \bibnamefont {Kimura}},\ }\bibfield  {title} {\bibinfo {title} {Generating radially polarized beams interferometrically},\ }\href {https://doi.org/10.1364/AO.29.002234} {\bibfield  {journal} {\bibinfo  {journal} {Appl. Opt.}\ }\textbf {\bibinfo {volume} {29}},\ \bibinfo {pages} {2234} (\bibinfo {year} {1990})}\BibitemShut {NoStop}%
\bibitem [{\citenamefont {Volke-Sepulveda}\ and\ \citenamefont {Ley-Koo}(2006)}]{VolkeSepulveda2006}%
  \BibitemOpen
  \bibfield  {author} {\bibinfo {author} {\bibfnamefont {K.}~\bibnamefont {Volke-Sepulveda}}\ and\ \bibinfo {author} {\bibfnamefont {E.}~\bibnamefont {Ley-Koo}},\ }\bibfield  {title} {\bibinfo {title} {General construction and connections of vector propagation invariant optical fields: {TE} and {TM} modes and polarization states},\ }\href {https://doi.org/10.1088/1464-4258/8/10/008} {\bibfield  {journal} {\bibinfo  {journal} {J. Opt. A: Pure Appl. Opt.}\ }\textbf {\bibinfo {volume} {8}},\ \bibinfo {pages} {867} (\bibinfo {year} {2006})}\BibitemShut {NoStop}%
\bibitem [{\citenamefont {Arfken}\ \emph {et~al.}(2011)\citenamefont {Arfken}, \citenamefont {Weber},\ and\ \citenamefont {Harris}}]{Arfken2011}%
  \BibitemOpen
  \bibfield  {author} {\bibinfo {author} {\bibfnamefont {G.~B.}\ \bibnamefont {Arfken}}, \bibinfo {author} {\bibfnamefont {H.~J.}\ \bibnamefont {Weber}},\ and\ \bibinfo {author} {\bibfnamefont {F.~E.}\ \bibnamefont {Harris}},\ }\href@noop {} {\emph {\bibinfo {title} {Mathematical Methods for Physicists: A Comprehensive Guide}}}\ (\bibinfo  {publisher} {Academic Press},\ \bibinfo {address} {Amsterdam},\ \bibinfo {year} {2011})\BibitemShut {NoStop}%
\bibitem [{\citenamefont {Kythe}(2019)}]{Kythe2019}%
  \BibitemOpen
  \bibfield  {author} {\bibinfo {author} {\bibfnamefont {P.~K.}\ \bibnamefont {Kythe}},\ }\href {https://doi.org/10.1201/9781315180236} {\emph {\bibinfo {title} {Handbook of {Conformal Mappings} and {Applications}}}}\ (\bibinfo  {publisher} {Chapman and Hall/CRC},\ \bibinfo {address} {New York},\ \bibinfo {year} {2019})\BibitemShut {NoStop}%
\bibitem [{\citenamefont {Wunsch}(1983)}]{Wunsch1983}%
  \BibitemOpen
  \bibfield  {author} {\bibinfo {author} {\bibfnamefont {A.~D.}\ \bibnamefont {Wunsch}},\ }\href@noop {} {\emph {\bibinfo {title} {Complex Variables with Applications}}}\ (\bibinfo  {publisher} {Addison-Wesley},\ \bibinfo {address} {Reading, MA},\ \bibinfo {year} {1983})\BibitemShut {NoStop}%
\bibitem [{\citenamefont {Freund}(2001)}]{Freund2001}%
  \BibitemOpen
  \bibfield  {author} {\bibinfo {author} {\bibfnamefont {I.}~\bibnamefont {Freund}},\ }\bibfield  {title} {\bibinfo {title} {Poincar{\'e} vortices},\ }\href {https://doi.org/10.1364/OL.26.001996} {\bibfield  {journal} {\bibinfo  {journal} {Opt. Lett.}\ }\textbf {\bibinfo {volume} {26}},\ \bibinfo {pages} {1996} (\bibinfo {year} {2001})}\BibitemShut {NoStop}%
\bibitem [{\citenamefont {Dennis}(2002)}]{Dennis2002}%
  \BibitemOpen
  \bibfield  {author} {\bibinfo {author} {\bibfnamefont {M.~R.}\ \bibnamefont {Dennis}},\ }\bibfield  {title} {\bibinfo {title} {Polarization singularities in paraxial vector fields: morphology and statistics},\ }\href {https://doi.org/10.1016/S0030-4018(02)02088-6} {\bibfield  {journal} {\bibinfo  {journal} {Opt. Commun.}\ }\textbf {\bibinfo {volume} {213}},\ \bibinfo {pages} {201} (\bibinfo {year} {2002})}\BibitemShut {NoStop}%
\bibitem [{\citenamefont {Otte}\ \emph {et~al.}(2016)\citenamefont {Otte}, \citenamefont {Alpmann},\ and\ \citenamefont {Denz}}]{Otte2016}%
  \BibitemOpen
  \bibfield  {author} {\bibinfo {author} {\bibfnamefont {E.}~\bibnamefont {Otte}}, \bibinfo {author} {\bibfnamefont {C.}~\bibnamefont {Alpmann}},\ and\ \bibinfo {author} {\bibfnamefont {C.}~\bibnamefont {Denz}},\ }\bibfield  {title} {\bibinfo {title} {Higher-order polarization singularities in tailored vector beams},\ }\href {https://doi.org/10.1088/2040-8978/18/7/074012} {\bibfield  {journal} {\bibinfo  {journal} {J. Opt.}\ }\textbf {\bibinfo {volume} {18}},\ \bibinfo {pages} {074012} (\bibinfo {year} {2016})}\BibitemShut {NoStop}%
\bibitem [{\citenamefont {Goodman}(2017)}]{Goodman2017}%
  \BibitemOpen
  \bibfield  {author} {\bibinfo {author} {\bibfnamefont {J.~W.}\ \bibnamefont {Goodman}},\ }\href@noop {} {\emph {\bibinfo {title} {Introduction to {Fourier} Optics}}},\ \bibinfo {edition} {4th}\ ed.\ (\bibinfo  {publisher} {W.~H. Freeman},\ \bibinfo {address} {New York},\ \bibinfo {year} {2017})\BibitemShut {NoStop}%
\bibitem [{\citenamefont {Goldstein}(2011)}]{Goldstein2011}%
  \BibitemOpen
  \bibfield  {author} {\bibinfo {author} {\bibfnamefont {D.~H.}\ \bibnamefont {Goldstein}},\ }\href@noop {} {\emph {\bibinfo {title} {Polarized Light}}}\ (\bibinfo  {publisher} {CRC Press},\ \bibinfo {address} {Boca Raton},\ \bibinfo {year} {2011})\BibitemShut {NoStop}%
\bibitem [{\citenamefont {Cox}\ and\ \citenamefont {Rosales-Guzm{\'a}n}(2023)}]{Cox2023}%
  \BibitemOpen
  \bibfield  {author} {\bibinfo {author} {\bibfnamefont {M.~A.}\ \bibnamefont {Cox}}\ and\ \bibinfo {author} {\bibfnamefont {C.}~\bibnamefont {Rosales-Guzm{\'a}n}},\ }\bibfield  {title} {\bibinfo {title} {Real-time {Stokes} polarimetry using a polarization camera},\ }\href {https://doi.org/10.1364/AO.485855} {\bibfield  {journal} {\bibinfo  {journal} {Appl. Opt.}\ }\textbf {\bibinfo {volume} {62}},\ \bibinfo {pages} {7828} (\bibinfo {year} {2023})}\BibitemShut {NoStop}%
\bibitem [{\citenamefont {Ndagano}\ \emph {et~al.}(2016)\citenamefont {Ndagano}, \citenamefont {Sroor}, \citenamefont {McLaren}, \citenamefont {Rosales-Guzm\'{a}n},\ and\ \citenamefont {Forbes}}]{Ndagano2016}%
  \BibitemOpen
  \bibfield  {author} {\bibinfo {author} {\bibfnamefont {B.}~\bibnamefont {Ndagano}}, \bibinfo {author} {\bibfnamefont {H.}~\bibnamefont {Sroor}}, \bibinfo {author} {\bibfnamefont {M.}~\bibnamefont {McLaren}}, \bibinfo {author} {\bibfnamefont {C.}~\bibnamefont {Rosales-Guzm\'{a}n}},\ and\ \bibinfo {author} {\bibfnamefont {A.}~\bibnamefont {Forbes}},\ }\bibfield  {title} {\bibinfo {title} {Beam quality measure for vector beams},\ }\href {https://doi.org/10.1364/OL.41.003407} {\bibfield  {journal} {\bibinfo  {journal} {Opt. Lett.}\ }\textbf {\bibinfo {volume} {41}},\ \bibinfo {pages} {3407} (\bibinfo {year} {2016})}\BibitemShut {NoStop}%
\end{thebibliography}

%

\end{document}